\documentclass[epj]{svjour}
\usepackage{graphicx,color,epsfig,rotate}

\begin{document}
\title{Thermodynamic studies of the two dimensional Falicov-Kimball model on a triangular lattice}
\author{Umesh K. Yadav \and T. Maitra \and Ishwar Singh}
%
%
\institute{Department of Physics, Indian Institute of Technology Roorkee,
Roorkee- 247667, Uttarakhand, India}
%
%
\abstract{
Thermodynamic properties of the spinless Falicov-Kimball model are studied on a triangular lattice using numerical
diagonalization technique with Monte-Carlo simulation algorithm. Discontinuous metal-insulator transition
is observed at finite temperature. Unlike the case of square lattice, here we
observe that the finite temperature effect is not able to smear out the
discontinuous metal-insulator transition seen in the ground state. Calculation of specific
heat ($C_v$) shows single and double peak structures for different values of
parameters like on-site correlation strength ($U$), $f-$electron energy
($E_f$) and temperature.
} 
\maketitle
\section{Introduction}
\label{intro}
The Falicov-Kimball model (FKM)~\cite{fkm} is one of the simplest and
most successful model for the correlated electron systems. This model was introduced
to describe the semiconductor-metal transition in mixed-valence compou-\hskip 3cm nds and
rare-earth systems. Recently systems such as tran-\hskip 3cm sition-metal dichalcogenides~
\cite{aebi,qian2,cava}, cobaltates~\cite{qian}, $GdI_{2}$~\cite{gd1,gd2} and
its doped variant $GdI_{2}H_{x}$~\cite{gd3} have attracted considerable
attention as they exhibit remarkable cooperative phenomena, like metal-
insulator transition, charge and magnetic order, excitonic instability
~\cite{aebi} and possible non-Fermi liquid states~\cite{gd2,castro}. These
systems are characterized by the presence of localized and itinerant electrons
confined to the two-dimensional triangular lattice.

In the FKM effective interactions are mediated by band electrons~\cite{kennedy,lieb}.
Therefore, the model has also been studied as a model of crystallization,
for example, as a model of binary alloy. The model describes the systems
exhibiting different types of phase configurations e.g. regular phas-\hskip 3cm e~\cite{maska}, phase
separation~\cite{free_leman,free_lieb,letf_free,free_grub,farkos}
and stripe phases~\cite{leman_free,leman_banach,haller_kennedy}. There
are predominantly more rigorous results available for the FKM on a bipartite lattice than
on non-bipartite lattices. On a bipartite lattice, one of the most important
result proved by Kennedy and Lieb~\cite{kennedy,lieb} is that at low enough
temperature the half-filled FKM possesses a long range order, i.e., the
localized electrons form a checkerboard pattern, the same as in the ground
state. This result holds for arbitrary bipartite lattices in dimensions $d \geq 2$
and for all values of on-site Coulomb correlation strength $U$.

\begin{figure*}[ht]
\begin{center}
\includegraphics[trim = 0.5mm 0.5mm 0.5mm 0.5mm, clip,width=12.0cm]{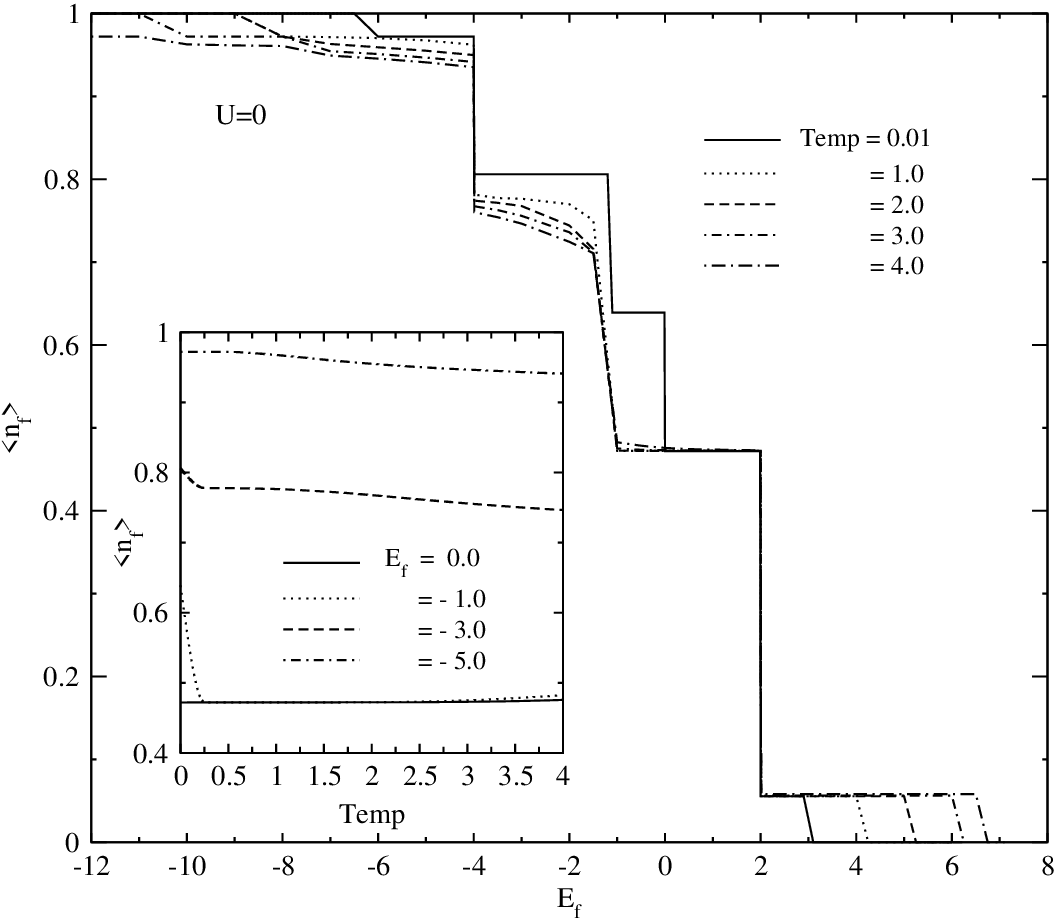}
\caption{($n_f$ - $E_f$) phase diagram for different values of temperature at U=0 and
temperature dependence of the f-electron occupation number $n_{f}$ at $U=0$ for different values of $E_f$ (shown in inset).}
\end{center}
\end{figure*}

Recently few interesting results are reported for the phase diagram of
localized $f-$electrons due to the inclusion of correlated hopping
($t^{\prime}$)~\cite{wojt,hirsch2,bulk,shv,farkov,farkov_squ,umesh1} of
itinerant electrons in FKM. Phase segregation, even in weak correlation
limit ($U \sim 0.5$ or $1$), is one of the most striking outcome of the
generalization of the FKM on bipartite~\cite{free_lieb,farkov_squ,kennedy_seg}
and on non-bipartite~\cite{umesh1} lattices.

Even though the FKM is one of the simplest models for correlated electrons,
a clear picture of metal-insulator transition in FKM is still debated for
many years. Metal-insulator transition strongly depends upon the choice of
the d-electron density of states (DOS). There are only few exact results
available on metal-insulator transition on a non-bipartite lattice~\cite{umesh1,maska_tri,umesh2}.
In a recent work~\cite{umesh1} we have studied the ground
state properties of the Falicov-Kimball Model on a triangular lattice with
correlated hopping and observed first order phase transition in the $f(d)-$
electron occupation at particular value of $f-$electron energy $E_f$. This
shows that the small change in the lattice parameter by means of introducing
impurity or by applying pressure one can successfully explain the valence and
metal - insulator transitions. In a separate work~\cite{umesh2} we
studied the effects of Coulomb repulsion between f- and d-electrons (U) and
also between f-electrons($U_f$) themselves in a two-fold degenerate f-level on
the ground state properties. We observed first order (discontinuous) insulator-
metal transition at particular values of $U$ and $U_f$.

A very important question, in this context arises as to what is the nature of
these phase-transitions at finite temperature? Does this model again give
the first order (discontinuous) insulator to metal transitions of f(d)-
electron occupation with $E_f$ and temperature? In addition it is also
interesting to know the dependence of the specific heat ($C_v$) on temperature
for different range of parameter values.

In this work, therefore, we study the FKM in its spinless version numerically for all
ranges of interactions, and explore the metal-insulator transition and specific heat on
a triangular lattice at finite temperature.

The Hamiltonian of the system may be written as
\begin{eqnarray}
{H} =-\sum_{\langle ij\rangle}
(t_{ij}+\mu\delta_{ij})d^{\dagger}_{i}d_{j}
\nonumber\\
+E_f \sum_{i} f^{\dagger}_{i}f_{i}
+U \sum_{i}f^{\dagger}_{i}f_{i}d^{\dagger}_{i}d_{i}.
\end{eqnarray}
\noindent Here $d^{\dagger}_{i}, d_{i}\,(f^{\dagger}_{i}, f_{i}$)  are, respectively,
the creation and annihilation operators for itinerant $d$-electrons (localized
$f$-electrons) at the site $i$. The first term in Eq.(1) is a measure of kinetic energy
of $d$-electrons on a triangular lattice: only nearest-neighbor hopping is
considered. The second term represents the dispersionless energy level $E_{f}$
of the $f$-electrons while the third term is the on-site Coulomb repulsion
between $d$- and $f$-electrons.

\section{Methodology}

In the Hamiltonian (1), hybridization between d- and f-electrons is absent.
Therefore, local $f$-electron occupation number $\hat{n}_{f,i}=f^{\dagger}_{i}f_{i}$, is
conserved. The $\omega_{i}=f^{+}_{i}f_{i}$
is a good quantum number as $[\hat{n}_{f,i},H]=0$ taking values either 1 or 0, according to the
whether site $i$ is occupied or unoccupied by an f-electron. Due to the local conservation
of $f$-electron number the Hamiltonian (1) may be written as,
\begin{equation}
H=\sum_{\langle ij\rangle}h_{ij}(\omega)d^{+}_{i}d_{j}+E_f\sum_{i}\omega_{i}
\end{equation}
\noindent where $h_{ij}(\omega)=-t_{ij}+(U\omega_{i}-\mu)\delta_{ij}$.

We set the hopping integral $t_{ij}=1$, for nearest neighboring sites
$i$ and $j$ and $t_{ij}=0$, otherwise. The eigenvalue spectrum of the Hamiltonian
$H$, for a configuration $\omega$ of $f$-electrons is calculated by numerical
diagonalization on a triangular lattice of finite size with periodic boundary
conditions (PBC). The average value of physical quantities is obtained by the
classical Monte Carlo method using Metropolis algorithm. The details of the
method can be found in our earlier papers~\cite{umesh1,umesh2}.

\begin{figure}[ht]
\begin{center}
\includegraphics[trim = 0.5mm 0.5mm 0.5mm 0.5mm, clip,width=7.6cm]{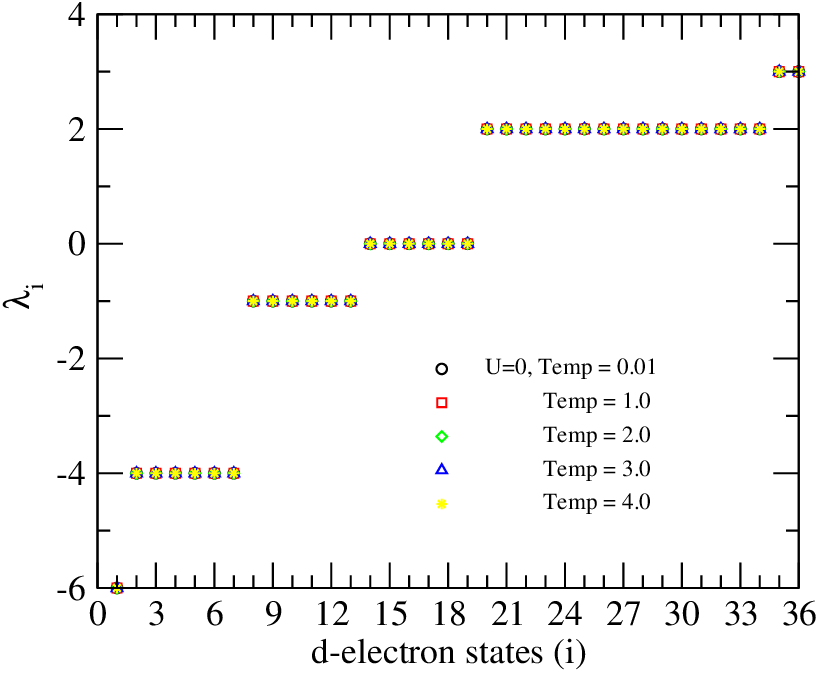}
\caption{(color online) The d-electron energy spectrum for U=0.}
\end{center}
\end{figure}

We work at half-filling, i.e., $N_{f}+N_{d}=N$ where $N_{f},\, N_{d}$ are the total
number of $f$- and $d$-electrons and $N=L^{2}~(L=6)$ is the total number of sites.
At a finite temperature T, the thermodynamic quantities are determined as averages
over various configurations $\omega(N_{f})$ with statistical weight P($\omega(N_{f})$) given
by
\begin{eqnarray}
{P(\omega(N_{f}))}=\frac{e^{-{\beta} F(\omega(N_{f}))}}{\cal Z}
\end{eqnarray}
\noindent where the corresponding free energy is given as,
\begin{eqnarray}
{F(\omega)}=-\,\frac{1}{\beta}\,[ln(\,\prod_{i}\,e^{-\beta E_{f}\omega_{i}})
\nonumber\\
+\sum_{j}\,ln(e^{-[\lambda_{j}(\omega(N_{f}))-\mu]\beta}+1)],
\end{eqnarray}
with $\mu$ being the chemical potential. The $\mu$ has been fixed to satisfy the condition
$n_f$+$n_d$=1.

\noindent The partition function is
\begin{eqnarray}
{\cal Z}=\prod_{i}(\sum_{\omega_i=0,1}e^{-\beta E_{f}\omega_{i}})\,\prod_{j}^{N}
(1+e^{-\beta[\lambda_{j}(\omega(N_{f}))-\mu]}).
\end{eqnarray}

\noindent Let a thermodynamic quantity `A' have value A($\omega(N_{f})$) corresponding to
the configuration $\omega(N_{f})$ of $N_{f}$-electrons, then the ensemble average of `A'
at temperature $T$ is obtained as
\begin{eqnarray}
{\langle A \rangle}=\,\frac{{\sum_{N_{f}}}{\sum_{\omega}}\,A(\omega(N_{f}))\,e^{-{\beta F(\omega(N_{f}))}}}{\cal Z}
\end{eqnarray}

\noindent For example the ensemble average of number of f-electrons for given values of U, E$_{f}$,
T is obtained as
\begin{eqnarray}
{\langle N_{f} \rangle}=\frac{{\sum_{N_{f}}}\,{\sum_{\omega}}\, N_{f}(\omega(N_{f})) \, e^{-{\beta F(\omega(N_{f}))}}}{\cal{Z}}
\end{eqnarray}

\noindent and $n_{f}=\langle n_{f} \rangle =\frac{\langle N_{f} \rangle}{N}$.

\noindent The total internal energy $E(\omega)$ of the system corresponding to
the configuration $\omega(N_{f})$ is given as,
\begin{eqnarray}
{E(\omega(N_{f}))}=\sum_{i}\,\frac{\lambda_{i}(\omega)}{e^{(\lambda_{i}(\omega)-\mu)\beta}+1}+\,E_{f}N_{f}(\omega)
\end{eqnarray}

\begin{figure}[ht]
\begin{center}
\includegraphics[trim = 0.5mm 0.5mm 0.5mm 0.5mm, clip,width=7.6cm]{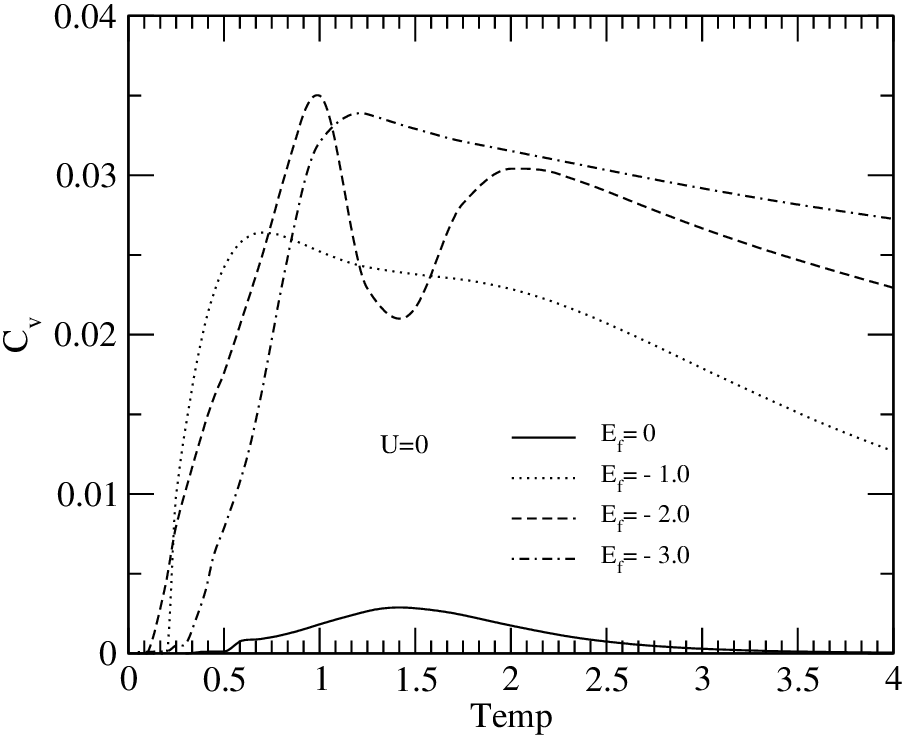}
\caption{Specific heat ($C_v$) as a function of temperature calculated at $U=0$ for different values of $E_f$.}
\end{center}
\end{figure}

The specific heat ($C_v$) is calculated from the fluctuation-dissipation theorem (FDT). The specific
heat is related through the FDT to the internal energy and defined as,
\begin{eqnarray}
{C_{v}}=\frac{1}{N}\,\frac{1}{kT^2}(\langle E^2 \rangle - {\langle E \rangle}^{2}),
\end{eqnarray}

\section{Results and discussion}

\subsection{Noninteracting case (U=0)}

\begin{figure*}[ht]
\begin{center}
\includegraphics[trim = 0.5mm 0.5mm 0.5mm 0.5mm, clip,width=14.0cm]{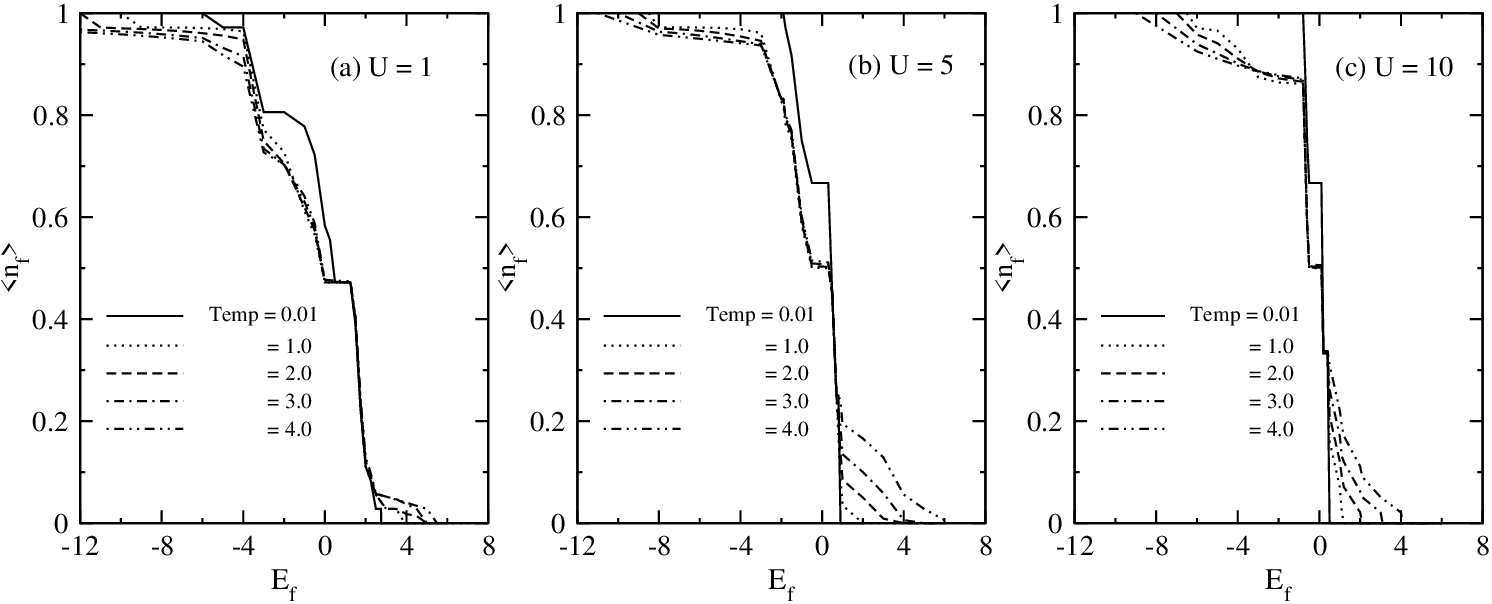}
\caption{($n_f$ - $E_f)$ phase diagram for different values of temperature calculated at (a) $U=1$, (b) $U=5$ and (c) $U=10$.}
\end{center}
\end{figure*}
\begin{figure*}[ht]
\begin{center}
\includegraphics[trim = 0.5mm 0.5mm 0.5mm 0.5mm, clip,width=14.0cm]{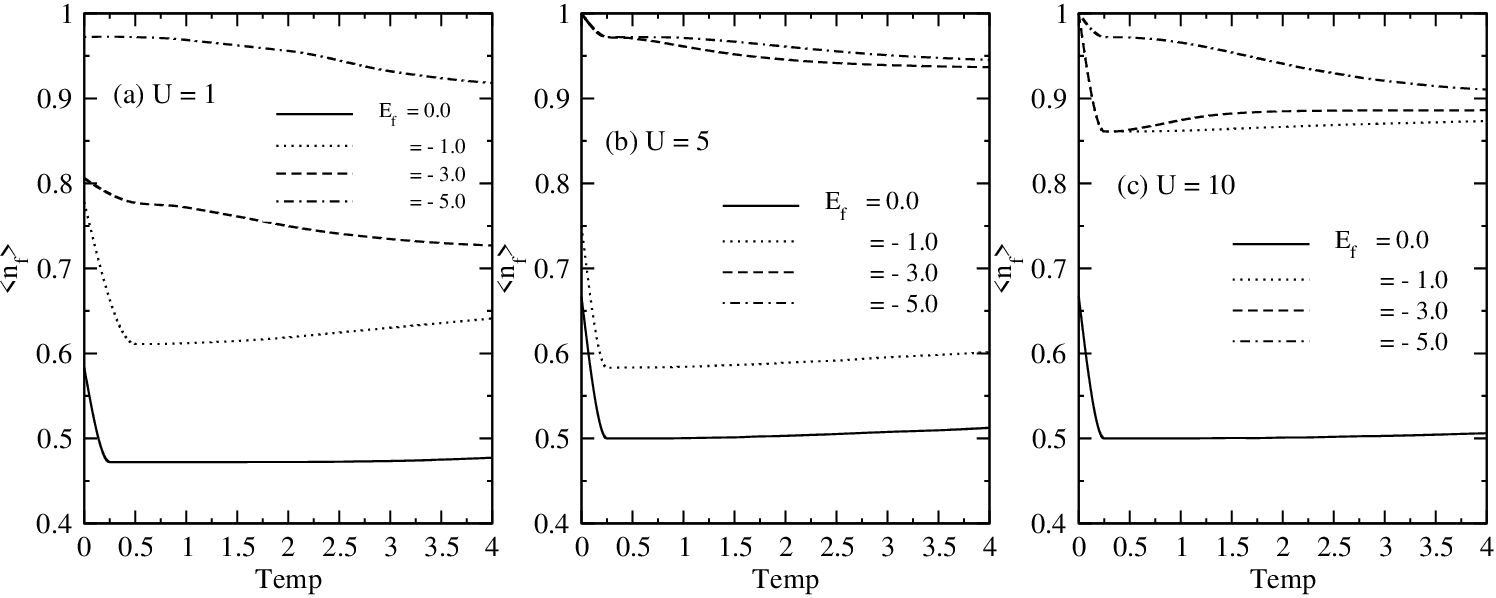}
\caption{Temperature dependence of the f-electron occupation number $n_{f}$ at (a) $U=1$, (b) $U=5$ and (c) $U=10$
for different values of $E_f$.}
\end{center}
\end{figure*}

We firstly study the noninteracting case ($U=0$). Using the method described
above, we have found average value of f-electron occupation number per site
$n_{f}=\langle n_f \rangle = \frac{\langle N_{f} \rangle}{N}$ as a function of
$E_f$ at different temperatures. Also we have calculated specific heat $C_v$
as a function of temperature for different values of $E_f$. In Fig.$1$ we have
shown $n_f$ as a function of $E_f$ for different temperatures and also as a
function of temperature for different values of $E_f$ (shown in inset of
Fig.$1$). In Fig.$2$ we show the d-electron energy spectrum corresponding to the configuration $\omega(N_{f})$
with minimum internal energy.

When $E_f$ is changed upwards (starting from the bottom of conduction band),
$n_f$ is found to decrease suddenly at some particular positions of $E_f$. For
example at $E_f$=-4.0, $n_f$ decreases abruptly from $n_f \sim 0.98$ to $n_f$=
0.8, at $E_f$=-1.0, from 0.8 to 0.6 at $E_f$=0.0, from 0.6 to $\sim$ 0.48 and
at $E_f$=2.0, from 0.48 to $\sim$ 0.03. If we look at the d-electron energy
spectrum, we find d-levels at energy E=-4.0, -1.0, 0.0, 2.0; i.e. there are
energy gaps~\cite{gap_defn} in d-electron spectrum in energy range [-4, -1],
[-1, 0], [0, 2]. As $n_f$+$n_d$=1, the chemical potential $\mu$ is pinned at
$E_f$. So, as $E_f$ is shifted upward, $\mu$ shifts upward. When  $\mu$ shifts
from -4 to -1, there being no d-levels in this range, $n_d$ remains constant
and so should $n_f$. This is exactly what we find as shown in Fig.$1$. As
$\mu$ reaches the energy -1, where there are many d-levels available, these
get occupied and $n_f$ decreases abruptly. Now with the increase of
temperature, d-states above chemical potential are also occupied fractionally.
That is why $n_f$ decreases a bit as temperature increases . However, in
contrast to the case of bipartite lattice~\cite{farkov_fintemp}, temperature
is unable to smear out the discontinuous transition on the triangular lattice.
Variation of $n_f$ with temperature at different $E_f$ is shown in the inset
of Fig.$1$. When $E_f$ is near the bottom of the d-band, $n_f$ is near unity at
$T\rightarrow0$ and decreases slowly as $T$ is increased. At other positions
of $E_f$, $n_f$ is lesser than unity (and the system is in classical mixed-
valent regime) and decreases slowly as $T$ is increased.

Shown in Fig.3 is the specific heat $C_v$ as a function of temperature for
different positions of $E_f$. A broad single peak is observed in specific heat
curve at all values of $E_f$ except at $E_f=-2.0$. At $E_f=-2.0$ there are two
peaks (one sharp peak followed by a broad peak). At low temperature we observe
that variation in $C_v$ deviates from the linear behavior of free electrons.
The nature of the peaks in the specific heat curves could be understood if we
compare these with the d-electron energy spectrum (shown in Fig.2) for
different temperatures. The d-electron spectrum does not depend on $N_f$,
$E_f$ and temperatures. The spectrum is discrete and multi-gaped. Wide gap
around Fermi energy in d-electron spectrum corresponds to broad peak and small
gap corresponds to sharp peak in $C_v$.

\subsection{Finite interaction case ($U \neq 0$)}

Let us consider the case where Coulomb correlation $U$ between d-and f-
electrons is finite. Fig.$4$(a), (b) and (c) show $(n_f-E_f)$ phase diagrams
for $U$=$1$, $5$ and $10$ and for temperatures $0.01$, $1$, $2$, $3$ and $4$
respectively. We observe that for U=$1$, as temperature increases from $0.01$ to $4$,
the valence transition (i.e. $n_f$-transition) (i) occurs at higher values of
$E_f$ and (ii) becomes smoother. We observe that the transition width (the
range of $E_{f}$ over which $n_{f}$ goes from $1$ to $0$), increases as
temperature increases from $0.01$ to $4$.

At U=$5$ and $10$ the $(n_f-E_f)$-transition at different temperatures shows
behavior similar to U=$1$ for $E_f$ well below or well above the middle of the
conduction band. When $E_f$ is near the middle of the conduction band, further
increment in the temperature does not affect the $n_f$-transition
significantly. This phenomena could be explained by the large gap in the many
body states around the Fermi energy ($E_F$). Due to large gap at the Fermi
energy (see Fig.$7$ (b) and (c) for U=$5$ and $10$ respectively) even higher
temperature is not sufficient to move the localized f-electrons to the higher
d-states.
%
\begin{figure*}[ht]
\begin{center}
\includegraphics[trim = 0.5mm 0.5mm 0.5mm 0.5mm, clip,width=12.0cm]{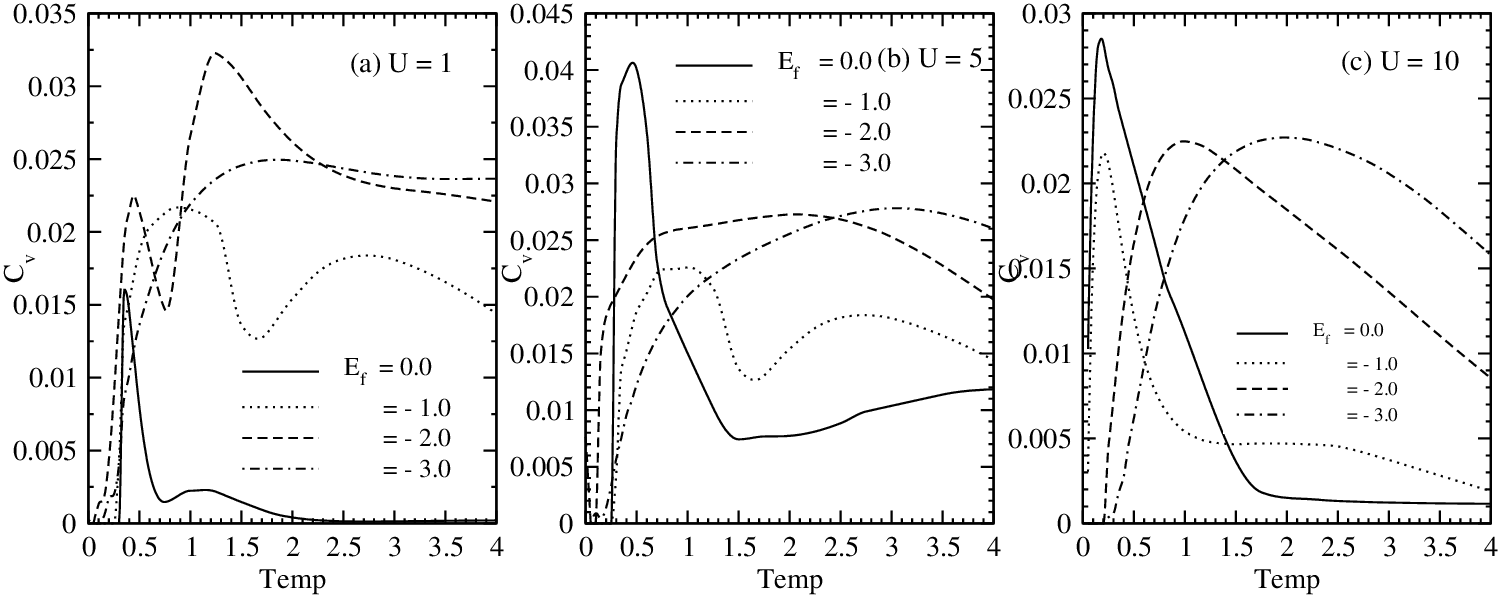}
\caption{Specific heat ($C_v$) as a function of temperature calculated at (a) $U=1$,(b) $U=5$ and (c) $U=10$ for different values of $E_f$.}
\end{center}
\end{figure*}

\begin{figure*}[ht]
\begin{center}
\includegraphics[trim = 0.5mm 0.5mm 0.5mm 0.5mm, clip,width=12.0cm]{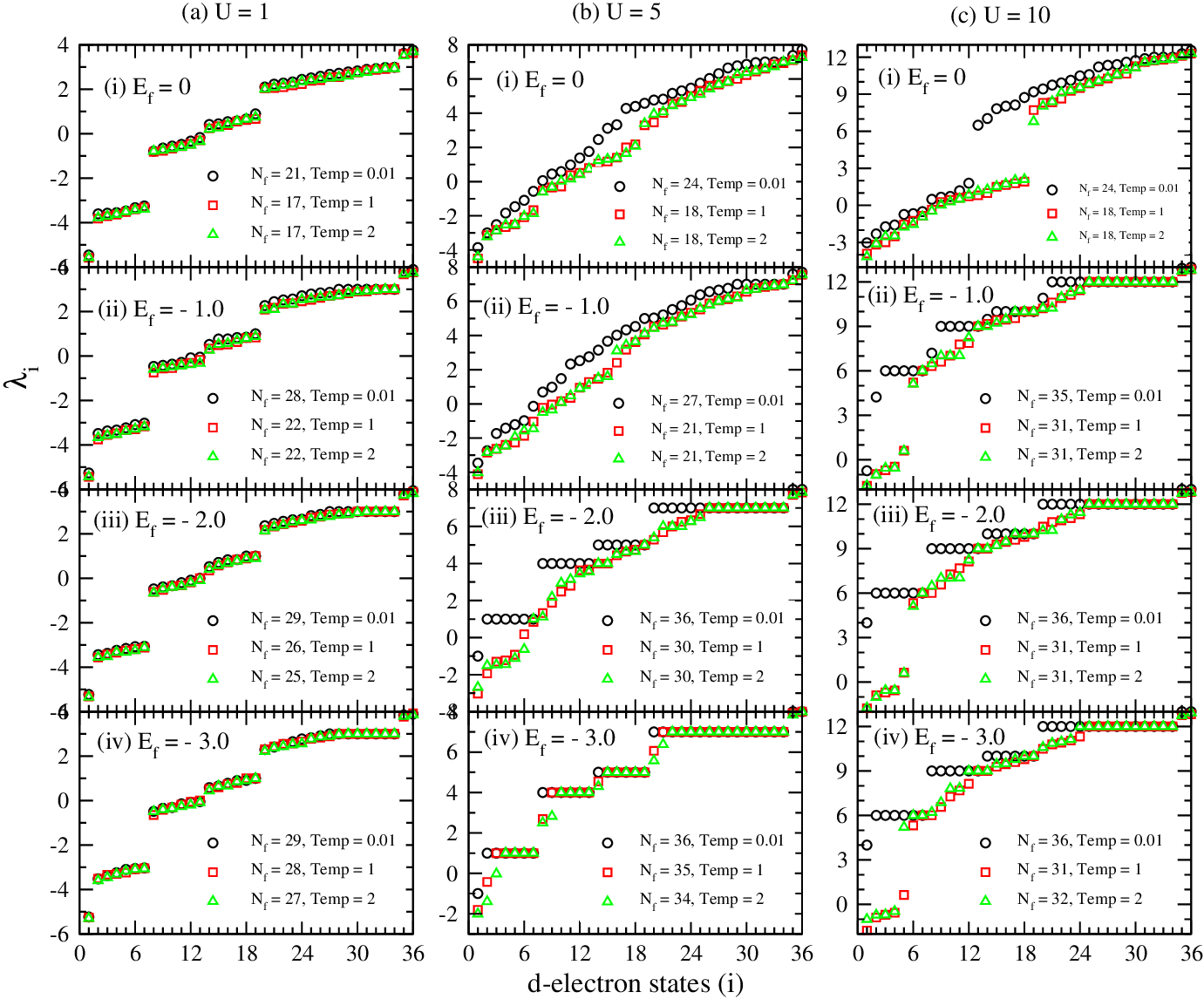}
\caption{(color online) The d-electron energy spectrum for different values of temperature and $E_f$ calcaluted at (a) U=1, (b) U=5 and (c) U=10.}
\end{center}
\end{figure*}

Fig.$5$(a), (b) and (c) show variation of $n_f$ with temperature for $U$=$1$,
$5$ and $10$ and for different $E_f$ positions. We observe that for all values
of $U$, $n_f$ approaches towards the value around 0.5 at high temperatures. At
low temperatures $n_f$ takes values in the range [0.5, 1.0] depending upon the
values of $U$ and $E_f$. Our results are in contrast to the results of P.
Farkasovsky~\cite{farkov_fintemp} obtained on bipartite lattice. We observed
discontinuous $(n_f-E_f)$ transitions with temperature. We report that even
higher temperature is not sufficient to smear the discontinuous $(n_f-E_f)$
transition observed in ground state spinless FKM on triangular lattice~\cite
{umesh1,umesh2}. They observed smooth variation in $n_f$ as a function of
temperature. The contrast between these two results are due to the choice of
different lattices. In square (bipartite) lattice, energy spectrum of the d-
electrons are symmetric about the middle of the conduction band, while in
triangular (non-bipartite) lattice the d-electron energy spectrum is
asymmetric about the middle of the conduction band and large number of d-
electron states lie in the small regime above the middle of the conduction band.

In Fig.$6$(a), (b) and (c) we have shown variation of $C_v$ with temperature
for $U$=$1$, $5$ and $10$ at different $E_f$ positions. In the small Coulomb
correlation limit (say $U$=1) and $E_f$ at the bottom of the conduction band
($E_f=-3.0$), a single broad peak in $C_v$ is observed. Moving the f-electron
energy level towards the middle of conduction band two peak pattern (one sharp
followed by a broad peak) is observed. For $E_f$ at center of conduction band
a sharp peak followed by a low broad peak is observed.
These different peaks can be explained by the energy spectrum of the d-electrons
shown in Fig.$7(a)$. In this figure we have shown the d-electron energy spectrum
corresponding to the configuration $\omega(N_{f})$ with minimum internal energy.
The d-electron spectrum is multi-gaped and depends upon the temperature and
$E_f$. For U=$1$, $E_f$=0 and for low temperature, a gapless spectrum is
observed. At finite temperature, we observe a gap around chemical potential in
d-electron spectrum. Moving $E_f$ from middle to bottom of conduction band say
$E_f=-2$, a large gap is observed at low temperature. The gap disappears at
higher temperatures. The small gap in d-electron spectrum leads to a sharp
peak while large gap corresponds to a broad peak in $C_v$.

For U=$5$ and $E_f$ at the bottom of conduction band a broad peak in $C_v$ is
observed. Width of peak is reduced by moving f-electron level towards middle
of conduction band. Two broad peaks in $C_v$ at $E_f=-1.0$ are observed. A
single peak is observed at $E_f$=0. At U=10, a single peak is observed in
$C_v$ for all observed $E_f$ values. The  width of the peak in $C_v$ decreases
with moving $E_f$ from bottom to middle of conduction band. The critical
temperature ($T_c$) at which peak occurs in $C_v$, reduces as $E_f$ moves from
bottom to middle of conduction band. If we compare $T_c$ for $E_f$=0 and at
different U-values we observe that $T_c$ goes on decreasing with increasing U,
similar to the result observed by M. Maska~\cite{maska_tri}. These features
are well explained on the basis of variation of d-electron spectrum with $E_f$
and temperature shown in Fig.7(b) and (c) for U=5 and 10 respectively.

In conclusion, we have studied the spinless Falicov-Kimball model on
a triangular lattice at finite temperature. The nature of metal-insulator
transition with temperature has been looked into in various parameter regime.
The temperature dependence of specific heat ($C_v$) is also studied for
different ranges of parameters like $U$,
$E_f$. Discontinuous metal-insulator transition is observed
even at finite temperature. This is unlike the case of bipartite lattices where
it has been reported that with temperature metal-insulator transition becomes
second order in nature. But here we see that the temperature can not smear the
discontinuous metal-insulator transition observed in ground state on triangular
lattice. Single and double peak patterns are observed in $C_v$ depending on the
parameters $U$, $E_f$ and temperature. It is proposed that the various features
could be explained by many body spectrum of $d-$electrons.

$Acknowledgments.$ UKY acknowledges CSIR, India for research fellowship.

\end{document}